# Impacts of Backside Insulation on the Dynamic On-Resistance of Lateral p-GaN HEMTs-on-Si

Yu-Xuan Wang, Mao-Chou Tai, Ting-Chang Chang*, Wei-Chen Huang, Zeyu Wan, Simon Li, Simon Sze, Guangrui Xia*

*Abstract*— We examined the effect of backside insulation on the dynamic on-resistance of lateral p-GaN HEMTs. To gain a comprehensive understanding of the dynamic on-resistance difference between substrate grounded and substrate floating p-GaN HEMTs, we conducted in-circuit double pulse testing and long-term direct current (DC) bias stress. We have realized that while backside insulation can enhance the breakdown voltage of lateral p-GaN HEMTs, it also comes with a tradeoff in device reliability. Results through Sentaurus TCAD simulation suggest that the use of backside insulation in devices gradually disperses potential to the buffer barrier. As a result, the potential barrier at the buffer edge of the 2DEG channel decreases significantly, leading to considerable electron trappings at buffer traps. This breakdown voltage and reliability tradeoff also applies to HEMT technologies using insulating substrates.

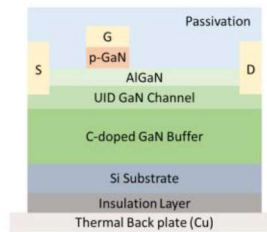

Fig. 1 Schematic diagram of the lateral p-GaN HEMT with backside insulation.

## I. INTRODUCTION

High electron mobility transistors (HEMTs) based on gallium nitride (GaN) have emerged as a leading technology in power electronics due to their exceptional performance, such as high breakdown, low on-resistance ($R_{on}$), and fast switching speed [1], [2]. These remarkable attributes have led to a growing demand for GaN HEMTs in high-power systems, such as renewable energy systems [3], electric vehicles [4], and biomedical applications [5]. However, ensuring reliable normally-off power switching devices is critical for safety. In terms of a positive threshold voltage ($V_{th}$), p-GaN HEMTs are regarded as the most favorable architecture among other GaN HEMT schemes. Therefore, p-GaN HEMTs on Si, which offer advantages such as low cost, large wafer size, and CMOS compatibility, are currently the most widely used structure. Despite that, p-GaN HEMTs on Si have still limited potential in higher voltage applications (typically >1.2 kV). This is due to the low critical electric field of Si materials (0.3

This work was supported in part by the National Science and Technology Council under Contract NSTC 112-2124-M-110-001 and Contract MOST 111-2221-E-A49-170-MY2. *(Corresponding authors: Ting-Chang Chang and Guangrui Xia)*

Y.-X. Wang and Simon Sze are with the Department of Electronics Engineering and Institute of Electronics, National Yang Ming Chiao Tung University, Hsinchu, 300, Taiwan

M.-C. Tai is with the Department of Photonics, National Sun Yat-Sen University, Kaohsiung, 80424, Taiwan

T.-C. Chang, and W.-C. Huang are with the Department of Physics, National Sun Yat-Sen University, Kaohsiung, 80424, Taiwan (email: tcchang3708@gmail.com)

Zeyu Wan and Gunagrui Xia are with the Department of Materials Engineering, The University of British Columbia (UBC), Vancouver, BC V6T 1Z4, Canada (email: gxia@mail.ubc.ca)

Simon Li is with the GaNPower International Incorporation, Burnaby, BC, V5M 2A4, Canada

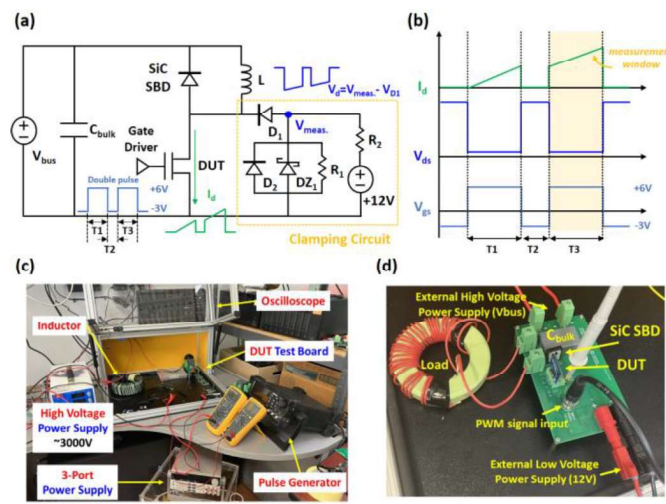

Fig. 2 (a) A schematic diagram of the switching circuit for double pulse testing. A clamping circuit is applied for ease of observation of the drain voltage. (b) Waveforms of gate voltage ($V_g$) and drain voltage ($V_d$) and current ($I_d$) for dynamic $R_{on}$ measurement. (c) Switching circuit setup for dynamic $R_{on}$ measurement. (d) An image of the PCB test board and its connection.

MV/cm [6]) and defects in the carbon-doped GaN buffer layer that results in vertical breakdowns [7], [8]. Therefore, to address this issue, much research has demonstrated characteristic improvements by transferring the substrate of p-GaN HEMTs from Si to other more insulating or wider band gap materials including sapphire [9], silicon carbide (SiC) [10], free-standing GaN [11], Qromis Substrate Technology (QST®) [12], silicon-on-insulator (SOI) [13], etc. The primary concept behind these approaches is to utilize the backside insulating layer to share the vertical voltage drop, thereby diminishing the electric field in the GaN buffer layer. While these approaches have succeeded in boosting the breakdown voltage ($V_{BR}$) of p-GaN HEMTs, the introduction of a more resistive substrate or underlayer could potentially affect the dynamic performance, which hasn't received a thorough examination

In this work, we investigated the impact of a backside insulation layer on the reliability of lateral p-GaN HEMTs. The reliability focuses on double pulse testing (DPT) and utilizes long-term direct current (DC) bias stress and Sentaurus TCAD to examine their differences. For those with backside insulation,

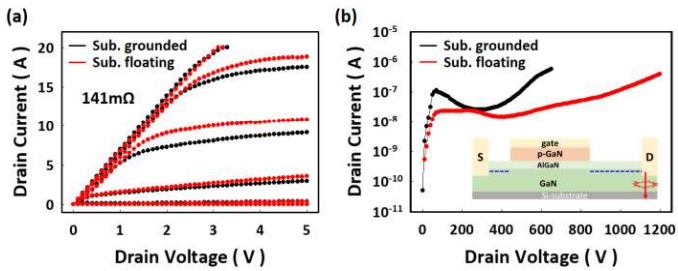

Fig. 3 (a) Output characteristics ($I_d$-$V_d$) and (b) off-state breakdown characteristics of the substrate grounded and the substrate floating p-GaN HEMT samples. Inset of Fig. 3 (b) is the schematic of vertical breakdown in HEMT devices.

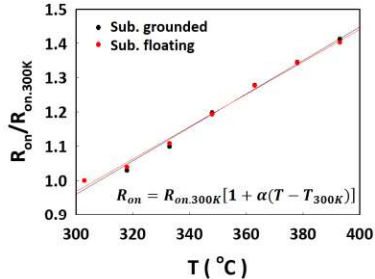

Fig. 4 The temperature dependence of on-resistance of both samples.

pronounced current collapse phenomena occur after DPT. The results suggest that the presence of the backside insulating layer dramatically degrades the off-state reliability, due to a backside-induced potential lowering effect that causes electron trappings at the buffer layer.

## II. EXPERIMENT

All experimental p-GaN HEMTs are provided by GaNPower International Inc. in Vancouver Canada. Two different types of samples were tested. One contains an insulating layer on the wafer backside [14], while the second does not. To differentiate the samples of each type, the former and latter will be referred to as substrate floating and substrate grounded respectively in the following discussions. The structure schematic of the substrate floating sample is demonstrated in Fig. 1. Fig. 2 (a) illustrates the circuit schematic for the DPT. A clamping circuit is applied for ease of observation of the drain voltage ($V_d$) [15]. Fig. 2 (c) and (d) show the test setup and a detailed overview of the print circuit board (PCB) test board. The test setup consists of a waveform generator, an oscilloscope, a 3-port power supply, a high-voltage power supply, and the device under test (DUT) connected to the PCB board. For the DPT waveforms, the gate voltage ($V_g$) of the DUT is continuously switched twice from -3 V to 6 V. The 1st on time ($T_1$), off time ($T_2$), and 2nd on time ($T_3$), are set as 8 μs, 1 μs, and 8 μs, respectively. The $V_d$ of the DUT is synchronized to the $V_g$ switching, while the off-state $V_d$ is determined by the $V_{bus}$ (the high-voltage power supply). The waveforms of gate voltage ($V_g$), drain voltage ($V_d$), and drain current ($I_d$) in DPT for dynamic Ron measurement are illustrated in Fig. 2 (b). For DC testing, transfer and output characteristics of the HEMT samples were characterized by an Agilent B1505 semiconductor parameter analyzer. All DC tests were conducted with the probes directly connected to the external pads of each chip. The threshold voltage ($V_{th}$) is determined as the gate voltage when the drain current density reaches the value of 1mA/mm.

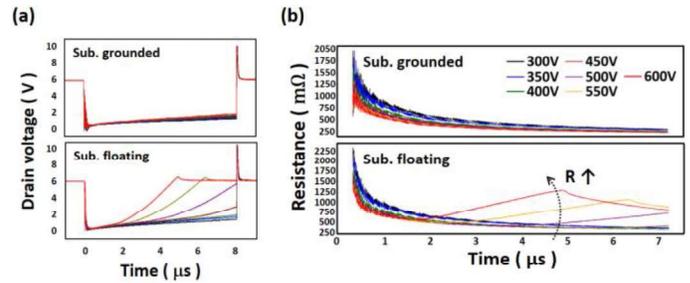

Fig. 5 (a) Drain voltage ($V_d$) and (b) dynamic $R_{on}$ of two HEMT samples with different drain bias stresses from 300 V to 600 V with a voltage step of 50 V. The $V_{g.on}$ is 6 V and $V_{g.off}$ is -3 V.

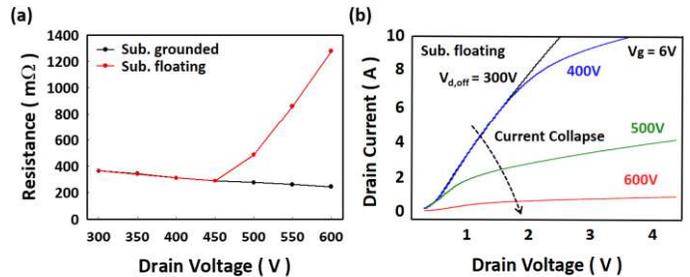

Fig. 6 (a) Dynamic $R_{ON}$ extracted at 450 ns of the second pulse. (b) Output characteristics ($I_d$-$V_d$) curves of the substrate floating HEMT after switching operations with different drain bias stresses. ($V_{d.off}$ = 300, 400, 500, 600 V)

## III. RESULT AND DISCUSSION

Fig. 3 (a) shows the output characteristics ($I_d$-$V_d$) of substrate grounded and substrate floating samples. The $R_{on}$ at $V_g$ = 6 V, evaluated from the slope at the linear region of the $I_d$-$V_d$, for both samples is approximately 141 mΩ. The off-state characteristics for both samples at $V_g$ = 0 V are presented in Fig. 3 (b). The substrate floating sample exhibits a $V_{BR}$ of 1270 V, which is much higher than that of the substrate grounded sample. As reported previously, the use of an insulating layer increases the $V_{BR}$ due to a reduction of the electrical field at the drain edge [13]. The floating substrate reduces the longitudinal electric field distribution on the GaN buffer, thereby further mitigating the substrate leakage current. As a result, the reduction in substrate leakage contributes to an increase in breakdown voltage. Notably, our backside design significantly enhances the breakdown characteristics, even with the insulating layer positioned behind the silicon substrate. In addition to the electrical behaviors, we also evaluate the thermal impact of our approach. Figure 4 illustrates the temperature dependencies of on-resistance ($R_{on}$) for both samples. The results indicate that the temperature dependencies remain consistent regardless of the insulating layer. Therefore, our design will not introduce additional heat-related issues.

Next, DPT was performed to investigate the switching behavior. Fig. 5 (b) demonstrates the time-dependent $R_{on}$ of both samples on the second pulse during the double pulse switch. The first 500 ns after switching is bypassed which is suggested as a measurement delay time that would cause misjudgments [16]. For the substrate grounded sample, a smooth decay of the $R_{on}$ through measuring time is observed which usually refers to electron detrappings [17]. On the other hand, for the substrate floating sample, a significant increase in $R_{on}$ is observed. To clarify the difference between the two

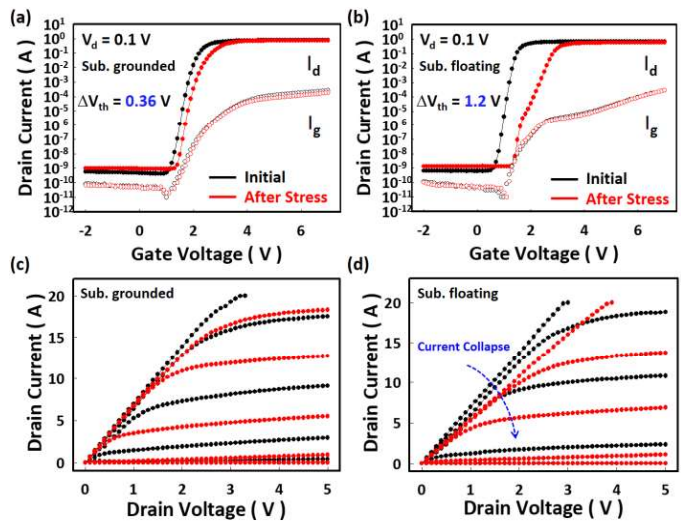

Fig. 7 $I_d$-$V_g$ and $I_g$-$V_g$ transfer characteristics of (a) the substrate grounded sample and (b) the substrate floating sample after off-state stress with $V_g$ = -3 V and $V_d$ = 500 V, respectively. $I_d$-$V_d$ output characteristics of (c) the substrate grounded sample and (b) the substrate floating sample after off-state stress. respectively.

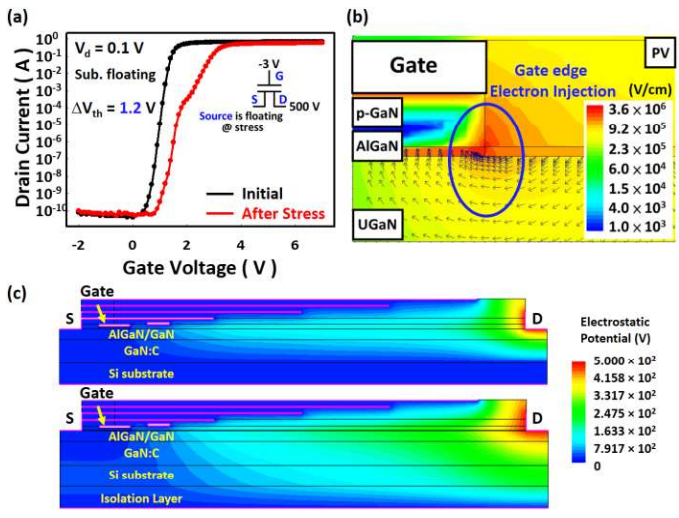

Fig. 8 (a) $I_d$-$V_g$ transfer characteristics of the substrate floating sample after off-state stress with $V_g$ = -3 V, $V_d$ = 500 V, while the source terminal remains floating during the stress. (b) The electric field distribution near the gate-drain edge of p-GaN HEMTs and (c) potential distribution profiles of the substrate grounded sample and the substrate floating sample under off-state reverse bias stress ($V_g$ = -3 V, $V_d$ = 500 V, and $V_s$ = 0 V).

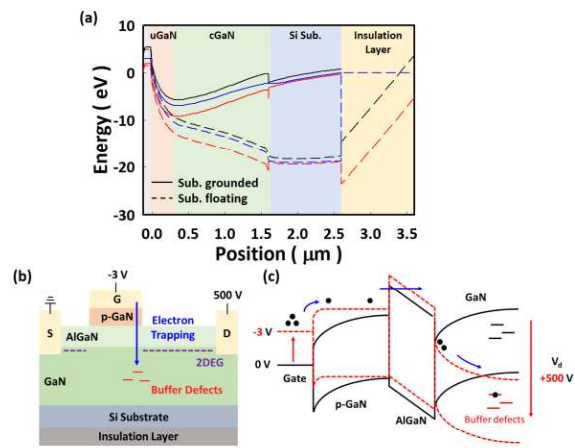

Fig. 9 (a) Vertical band diagram along the gate-drain edge of the substrate grounded sample and the substrate floating sample under off-state reverse bias stress ($V_g$ = -3 V, $V_d$ = 500 V, and $V_s$ = 0 V). (b) The schematic diagram and (c) the energy band diagram of electron trapping from gate to buffer defects in HEMTs under off-state stress.

samples, the $V_d$ of both samples on the second pulse are presented in Fig. 5 (a). The waveform of $V_d$ in substrate grounded sample indicates that DUT is operated in the linear region. However, an abnormal increase in $V_d$ is observed for the substrate floating sample compared to the substrate grounded sample. This phenomenon is attributed to the current collapse. During double pulse testing, inductance is employed to supply current to the DUT, thereby determining the $V_d$ of the DUT. However, a significant current collapse occurs when the sample utilizes the insulating design, leading to a reduced output current under the same gate voltage. Therefore, the voltage drop across the DUT is forced to increase to match the current flowing through it. This operation causes the DUT to operate in the saturation region rather than the linear region, consequently further increasing the measured resistance. Furthermore, Fig. 6 (b) shows the transfer $I_d$-$V_d$ curve of the substrate floating sample after different off-state switching biases. Accordingly, when increasing the off-state bias, the current collapse behavior gradually increases. Consequently, dynamic $R_{on}$ increases with the increasing drain bias when employing the insulating layer in the HEMTs, as shown in Fig 6 (a). On the other hand, a clamping circuit is integrated into the DPT circuit to facilitate the observation of $V_d$, limiting that the maximum measurable value of $V_d$ to 6 V. When the $V_d$ exceeds the threshold value, it is forced to reduce to 6 V. During DPT measurement with higher drain biases, such as 550 V and 600 V, the $V_d$ of the DUT needs to increase further due to significant current collapse. However, once $V_d$ exceeds 6 V, it is forced to drop back to 6 V due to the clamping circuit. Consequently, under higher drain bias conditions, such as 550 V and 600 V, the measured resistance experiences a sudden decrease thereafter, as depicted in Fig. 5 (b).

To focus on which duration of the pulse cycle leads to device degradation during DPT, long-term DC bias stresses were performed. Both on-state gate bias stress [18] (which resulted in negligible degradations and is therefore not shown) and off-state reverse bias stress (RBS) [19] were conducted. The RBS condition was given with $V_g$ = -3 V and $V_d$ = 500 V for 1000 s.

The $I_d$-$V_g$ transfer characteristics of both samples before and after stress are shown in Fig. 7 (a) and (b), respectively. Accordingly, a more severe $V_{th}$ shift for the substrate floating sample is observed. In addition, Fig. 7 (c) and (d) show the $I_d$-$V_d$ of both samples before and after stress, while current collapse is only observed in the substrate floating sample. The results in DC stress coincide with the DPT indicating that most of the degradation happens during the large $V_d$ duration ($T_2$).

Under RBS conditions, current collapse is commonly regarded as electron injection from the gate terminal [20]-[22]. This phenomenon can be confirmed through an identical RBS condition, but with a floating source terminal, as shown in Fig. 8 (a), where the degradation is similar regardless of the source connection. The distribution of electric field near the gate-drain edge under the given RBS condition is shown in Fig. 8 (b), which suggests that electrons would trap at either the virtual gate regions, AlGaN layer, or the buffer [20], [21]. Therefore, to confirm the electron trapping regions, the gate leakage ($I_g$) before and after stress for both samples were monitored, as shown in Fig. 7 (a) and Fig. 7 (b), respectively. Typically, electron trappings at either the virtual gate regions or the AlGaN layer will affect the $I_g$ [20], [21]. However, unaffected $I_g$ indicates that electron trappings specifically occur at the buffer layer, distinct from other regions. These defects either originate from the carbon-doped GaN buffer layer [23] or the dislocations from the substrate [24].

To elucidate the root cause of the pronounced trappings in the substrate floating HEMT, the Sentaurus TCAD was used to support our findings, particularly the potential distribution under the same RBS conditions, as shown in Fig. 8 (c). The noticeable gradual dispersion of the potential towards the insulating layer is evident in the comparison of both potential distributions. Consequently, the vertical band diagram along the gate-drain edge shown in Fig. 9 (a) demonstrates a reduction in the backside potential barrier of the 2DEG channel. The phenomenon of backside-induced potential lowering enhances the likelihood of electron trapping from the gate, leading to a shift in $V_{th}$ within the $I_d$-$V_g$ curve, accompanied by an elevation in $R_{on}$ and the occurrence of current collapse in the $I_d$-$V_d$ curve, as demonstrated in Fig. 7 (b) and Fig. 7 (d), respectively. The schematic of this mechanism and its corresponding band diagram are illustrated in Fig. 9 (b) and Fig. 9 (c).

## IV. CONCLUSION

In summary, our study explored the tradeoff between the high breakdown voltage ($V_{BR}$) and the reliability of lateral p-GaN HEMTs, that applies to HEMT technologies using insulating substrates and/or backside insulation layers. We found that the use of backside insulation can increase $V_{BR}$ by sharing the vertical potential, but it also decreases the buffer-edge potential barrier of the 2DEG channel, causing electron injections from the gate to become trapped at carbon-related traps or dislocations in the GaN buffer layer. This results in severe current collapse, not only undermining the conduction loss of the GaN devices but also further compromising the efficiency of high-power systems. Our findings suggest that other insulating substrate methods proposed in [9-13] may encounter similar challenges. An optimal balance between the $V_{BR}$ requirement and dynamic performance needs to be considered in the design stage.